\documentstyle[aps,prl,epsfig,twocolumn]{revtex}

\newcommand{\boldvec}[1]{\mbox{\boldmath$#1$}}

\newcommand{\annhilate}[2]{\hat{#1}^{\phantom\dag}_{{#2}}}
\newcommand{\create}[2]{\hat{#1}^{\dag}_{{#2}}}

\begin{document}

\draft

\title{Ramsey fringes in a Bose-Einstein condensate between atoms and
  molecules}

\author{S.J.J.M.F. Kokkelmans and M.J. Holland}

\address{JILA, University of Colorado and National Institute of
  Standards and Technology, Boulder, Colorado 80309-0440}

\wideabs{

\maketitle

\begin{abstract}
  In a recent experiment, a Feshbach scattering resonance was
  exploited to observe Ramsey fringes in a $^{85}$Rb Bose-Einstein
  condensate.  The oscillation frequency corresponded to the binding
  energy of the molecular state. We show that the observations are
  remarkably consistent with predictions of a resonance field theory
  in which the fringes arise from oscillations between atoms and
  molecules.
\end{abstract}


}

Interest in the physics of ultracold molecules has been growing
considerably in the past few years~\cite{wynar,heinzen}. One of the goals has 
been the creation of a molecular Bose-Einstein condensate (BEC). In a recent
experiment at JILA, performed by Donley {\it et al.}~\cite{donley}, a
coherence between atoms and molecules was demonstrated in a BEC of
$^{85}$Rb atoms. In this experiment, two magnetic field pulses were
applied to the condensate, and oscillations observed in the population
of two spatially distinguishable components as a function of the pulse
separation time.  One of the components, the `remnant' atoms, had a
similar spatial profile to the original BEC. This was in contrast to a
second component, the `burst' atoms, which moved away from the remnant
having gained a considerable amount of energy. In addition, there was
a missing fraction of atoms unaccounted for after the pulse sequence.
Significantly, the frequency of the population oscillations
corresponded to the binding energy of the highest lying molecular
state.

This observation followed other remarkable experiments carried out in
the same group. The collapse of a BEC was studied when a Feshbach
resonance was used to create a large negative scattering
length~\cite{donley1}.  Because of the rather violent destruction of
the collapsing condensate, this effect was dubbed a `Bosenova' in
analogy to a supernova explosion. In a precursor experiment to the one
we consider here, a single strong-coupling field pulse was applied,
and there remnant and burst atoms were also observed~\cite{claussen}.

In this letter, we describe the most recent JILA experiment by a
resonance field theory that has been developed over the last two
years~\cite{holland,holland2,kokkelmans}. This mean-field theory of
dilute atomic gases goes beyond the level of the Gross-Pitaevskii
equation~\cite{timmermans,mackie} to include the essential pairing physics 
necessary to
describe resonances in the two-body scattering. In the case of a Fermi
gas it was applied to describe superfluidity close to a Feshbach
resonance. Here, however, we are able for the first time to compare
this approach with experimental data, and we obtain remarkable
agreement with the observations.

A critical ingredient of the theory is the description of the Feshbach
resonance.  Such a resonance arises when a bound state (molecular
state) lies near the threshold of the collision continuum. This bound
state belongs effectively to a channel which is energetically closed.
In a system of two rubidium atoms, the position of this bound state
depends on magnetic field, due to hyperfine and Zeeman interactions.
The Feshbach resonance gives rise to a dispersive behavior of the
scattering length, that can be formulated accurately as

\begin{equation}
a(B)=a_{\rm bg}\left( 1-\frac{\Delta B}{B-B_0} \right).
\end{equation}
Here $a_{\rm bg}$ is the background scattering length, $\Delta B$ the
width of the resonance, and $B_0$ the field value where the scattering
length is infinity. The magnetic field can be easily converted into an
energy detuning of the molecular state from threshold by the relation
$ \nu_0 =(B - B_0) \Delta\mu$, where $\Delta\mu$ is the difference in
magnetic moments of the energetically open and closed channels. In our
theory, we treat the closed channel explicitly by using molecular
field operators and ascribing a coupling to the continuum for this
molecular state.

The Hamiltonian for the resonance system is given by
\begin{eqnarray}
\hat H &=& \int d^3 \boldvec x \left( \create{\psi}{a} 
(\boldvec x) H_a (\boldvec x)
\annhilate{\psi}{a}(\boldvec x)+ \create{\psi}{m} 
(\boldvec x) H_m (\boldvec x)
\annhilate{\psi}{m}(\boldvec x) \right) \nonumber \\
&+&\frac{1}{2} \int d^3 \boldvec x_1 d^3 \boldvec x_2 \Bigl[
\create{\psi}{a} (\boldvec
x_1) \create{\psi}{a} (\boldvec x_2) V(\boldvec x_1-\boldvec x_2)
\annhilate{\psi}{a} (\boldvec x_2) \annhilate{\psi}{a} 
(\boldvec x_1)  \nonumber\\
&+& \bigl(
\create{\psi}{m}(\frac{\boldvec x_1+\boldvec x_2}{2})
g(\boldvec x_1-\boldvec x_2) \annhilate{\psi}{a} (\boldvec x_2)
\annhilate{\psi}{a} (\boldvec x_1) +\rm{H.c.}\bigr) \Bigr] ,
\end{eqnarray}
where the field operators $\create{\psi}{a}(\boldvec x)$ and
$\create{\psi}{m}(\boldvec x)$ create an atom or molecule at position
$\boldvec x$, and H.c.\ denotes the Hermitian conjugate. The free
Hamiltonians $H_a(\boldvec x)=-\hbar^2\nabla^2/2m$ and $H_m(\boldvec
x)=-\hbar^2\nabla^2/4m+\nu$ for atoms with mass $m$ and molecules with
mass $2m$ include the detuning $\nu$.  Atom-molecule collisions and
molecule-molecule collisions give higher order corrections. The
potential terms $V(\boldvec x_1-\boldvec x_2)$ and $g(\boldvec
x_1-\boldvec x_2)$ have to be chosen such that both the scattering
physics and the molecular binding energies are correctly described. We
verify this by noting that the scattering equations are included in
this resonance mean-field theory~\cite{kokkelmans}. Although
unnecessary here, it is important from a fundamental perspective that
the above Hamiltonian can be further generalized by the inclusion of
more scattering resonances to systematically improve the description
of the two-body scattering physics to any desired precision.

Starting from this Hamiltonian we obtain the Hartree-Fock-Bogoliubov
(HFB) equations of motion.  We define expectation values for the
atomic and molecular condensates $\phi_a(\boldvec x)=\langle
\annhilate{\psi}{a} \rangle$ and $\phi_m(\boldvec x)=\langle
\annhilate{\psi}{m} \rangle$. Moreover, we define a $2 \times 2$
density matrix for the fluctuating components of the atomic field
operators $\annhilate{\chi}{a}=\annhilate{\psi}{a}- \langle
\annhilate{\psi}{a}\rangle$, that describe the noncondensed atoms:
\begin{equation}
 \cal G (\boldvec x,\boldvec y)= \left( \begin{array}{cc}
  \langle \create{\chi}{a}(\boldvec y)\annhilate{\chi}{a}
  (\boldvec x) \rangle  &
  \langle \annhilate{\chi}{a}(\boldvec y)\annhilate{\chi}{a}
  (\boldvec x) \rangle
  \\
  \langle \create{\chi}{a}(\boldvec y)\create{\chi}{a}
  (\boldvec x) \rangle  &
  \langle \annhilate{\chi}{a}(\boldvec y)\create{\chi}{a}
  (\boldvec x) \rangle
  \end{array} \right).
\end{equation}
The elements of this matrix can be given in terms of the normal
density $G_N (\boldvec x, \boldvec y)=\langle
\create{\chi}{a}(\boldvec y)\annhilate{\chi}{a}(\boldvec x) \rangle$
and the anomalous density $G_A (\boldvec x, \boldvec y)=\langle
\annhilate{\chi}{a}(\boldvec y) \annhilate{\chi}{a}(\boldvec x)
\rangle$~\cite{holland}.

To begin with, we consider a gas which is homogeneous and isotropic,
which results in a translationally invariant system. This implies that
the single particle fields are constant in space, and that the
two-particle fields depend on the coordinate difference $r=|\boldvec x
- \boldvec y|$ only. We substitute local interactions for the
potential terms: $V(r)=V \delta(r)$ and $g(r)=g \delta(r)$, where $V$
and $g$ are constants. The HFB equations for the density matrix $\cal
G$ and for the condensate fields $\phi_a$ and $\phi_m$ are obtained
from the Heisenberg equations for the field operators, having taken
expectation values and applied Wick's theorem:
\begin{eqnarray}
i\hbar\frac{d \phi_a}{dt}&=&V(|\phi_a|^2+2 G_N(0))\phi_a
       +(VG_A(0) + g\phi_m)\phi_a^*, \label{meanfieldat} \nonumber\\
          && \\
i\hbar\frac{d \phi_m}{dt}&=&\frac{g}{2}(\phi_a^2+G_A(0)) +\nu\phi_m, \\
i\hbar\frac{d G_N(r)}{dt}&=&2 {\rm Re}\left[
V(\phi_a^2+G_A(0))G_A^*(r)+g\phi_m G_A^*(r) \right], \\
i\hbar\frac{d G_A(r)}{dt}&=&-\frac{\hbar^2 \nabla^2}{2
\mu}G_A(r)
+4V(|\phi_a|^2 +G_N(0))G_A(r) \nonumber\\
&+& [V(\phi_a^2+G_A(0))+ g\phi_m](2G_N(r) + \delta(r)),
\label{meanfieldga}
\end{eqnarray}
with $\mu$ the reduced mass. This is the complete closed set of
equations to be dynamically solved. As emphasized earlier, the binary
collision physics encapsulated in the HFB equations is extracted by
setting the density dependent shifts to zero
\begin{eqnarray}
i\hbar \frac{d}{dt} {\cal P}(r)
&=&\left[ -\frac{\hbar^2 \nabla^2}{2 \mu}+V\delta(r) \right] {\cal P}(r)
+ g\delta(r)\phi_m, \\
i\hbar\frac{d}{dt} \phi_m &=& \frac{g}{2}
{\cal P}(0)+\nu\phi_m,
\end{eqnarray}
where ${\cal P}(r) =\phi_a^2+G_A(r)$ is the total pairing field,
leaving the two coupled two-body scattering equations with contact
interactions~\cite{kokkelmans}.

The local potentials in Eqns.~(\ref{meanfieldat})-(\ref{meanfieldga})
give rise to an ultraviolet divergence, which can be properly treated
by renormalization. This must be done in such a way as to maintain the
correct underlying two-body resonance physics for any momentum cutoff
$K$ in the field theory. One should consider the delta function
interactions to be appropriate zero range limits of nonlocal
potentials (e.g.\ square well potentials), and the properties of these
potentials can then be chosen such that the microscopic low energy
two-body physics around a Feshbach resonance is correctly described.
This renormalization procedure amounts to replacing the coupling
constants in the Hamiltonian by parameters which depend on $K$. The
required parameters can be concisely summarized by the complete
relations $V=\Gamma U$, $g=\Gamma g_0$, and $\nu= \nu_0 +\alpha g
g_0/2$, where $U=4\pi \hbar^2 a_{\rm bg}/m$, $\Gamma=(1-\alpha
U)^{-1}$, $\alpha=m K/(2\pi^2 \hbar^2)$, and $g_0$ is determined from
the field dependence of the binding energy as we now explain.
                                                 
\begin{figure}
\begin{center}\
  \epsfysize=60mm \epsfbox{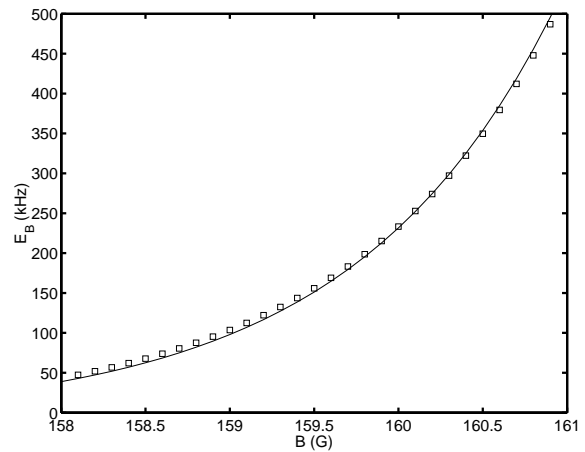}
\end{center}
\caption{
  Binding energies as a function of magnetic field. The solid line is
  the coupled channels result for the most accurate rubidium
  interactions.  This is compared with the binding energy resulting
  from a contact scattering model with $a_{\rm bg}=-450 a_0$ (where
  $a_0$ is the Bohr radius), $\Delta\mu=2.23\mu_B$ (with $\mu_B$
  the Bohr magneton), and $g_0=3.11\times 10^{-38}$~Jm$^{3/2}$
  (open squares).}
\label{fig1}
\end{figure}

A big advantage for this system is that the rubidium two-body
interactions are extremely well known. Typical scattering lengths can
be calculated to at least the 1\% level~\cite{kempen}. For this
experiment, where the fringe frequency is determined by the energy of
the highest bound state, accurate knowledge of the binding energy is
crucial. In Fig.~\ref{fig1} we show the binding energy as a function
of magnetic field using a full coupled channels
calculation~\cite{footnote}. This calculation consists of all relevant
rubidium two-body interactions including hyperfine and Zeeman
interactions and electron-spin dependent singlet and triplet
potentials. In order to reproduce exactly the result of the full
coupled-channels calculation, we would need to go beyond the single
resonance formulation we have presented.  However, in spite of the
complexity of the real rubidium system, the field theory with one
resonance state allows a remarkably good approximation to the binding
energy over the field range of interest.  This comparison is shown in
Fig.~\ref{fig1} and determines the value of $g_0$ used in our
following simulations.

\begin{figure}
\begin{center}\
  \epsfysize=60mm \epsfbox{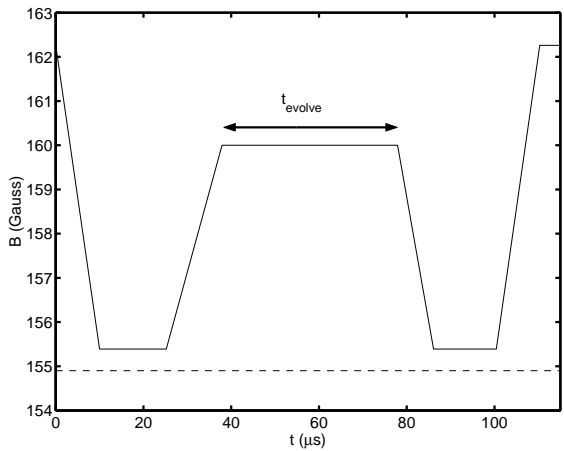}
\end{center}
\caption{
  A typical magnetic field pulse sequence as a function of time. The
  interval $t_{\rm evolve}$ is modified in the experiment. The
  position of the resonance is indicated by the dashed line.}
\label{fig2}
\end{figure}

The outcome of the experiment of Donley {\it et al.}~\cite{donley}
closely resembles the seminal experiments on Ramsey fringes in atomic
beam physics~\cite{ramsey}.  The starting point is a condensate of
$^{85}$Rb atoms in the $|f,m_f\rangle = |2,-2\rangle$ state, at a
magnetic field where the scattering length is close to zero. Two
magnetic field pulses are applied (a schematic is shown in
Fig.~\ref{fig2}), each of which brings the condensate close to
resonance. The two pulses are separated by a free evolution interval
$t_{\rm evolve}$, during which time the magnetic field is increased to
move the system further away from resonance.  After this pulse
sequence, the remaining number of atoms in the condensate is measured,
which is then called the remnant.  Also a burst of noncondensate atoms
is observed. The populations of the remnant and the burst both show
oscillations as function of $t_{\rm evolve}$ at a frequency that
corresponds to the binding energy of the molecular state at the
intermediate field.  The sum of the populations of the remnant and
burst do not add up to the initial number, implying a missing
component.  The theoretical solutions which follow in this Letter can
be directly compared with the results presented in the figures of
Donley {\it et al.}~\cite{donley}.

\begin{figure}
\begin{center}\
  \epsfysize=60mm \epsfbox{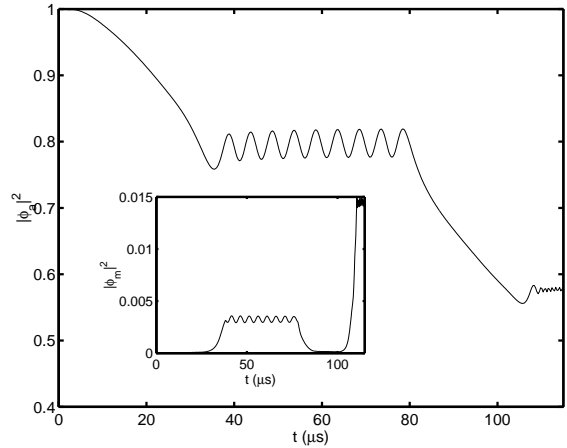}
\end{center}
\caption{
  Fraction of the atomic condensate as a function of time. We identify
  these atoms as the remnant of Donley {\it et
    al.}~\protect\cite{donley}. The calculation is done for the pulse
  sequence given in Fig.~\ref{fig2}, for a density $ n=3.9\times
  10^{12}$ cm$^{-3}$. The inset shows the analogous graph for the
  molecular condensate.}
\label{fig3}
\end{figure}

In spite of the fact that the mean-field
Eqns.~(\ref{meanfieldat})-(\ref{meanfieldga}) were derived for a
homogeneous system, we may apply our theory to a trapped gas. At the
highest energy scales reached during the evolution, the velocity of
the atoms is sufficient to move only one hundredth of the oscillator
ground state size in the tightest direction during the full evolution
time. Therefore, a complete quantum description of the oscillator
levels is not required. Instead, we use a local density approximation
and perform a Gaussian average over the densities of the gas. For each
value of the density, we solve the time-dependent equations modifying
the detuning according to the time-dependent field given in
Fig.~\ref{fig2}. In Fig.~\ref{fig3} we show a time evolution of the
atomic condensate, and the corresponding time evolution of the
molecular condensate. At the end of the first pulse about 25\% of the
condensate atoms have been converted into other components. The growth
of population in the molecular condensate takes place mostly during
the final ramp. It is notable that the fraction of molecules does not
account for the missing atoms.  In fact, the atoms are mainly
transfered to the normal and anomalous densities which are ascribed to
the noncondensate component.

\begin{figure}
\begin{center}\
  \epsfysize=60mm \epsfbox{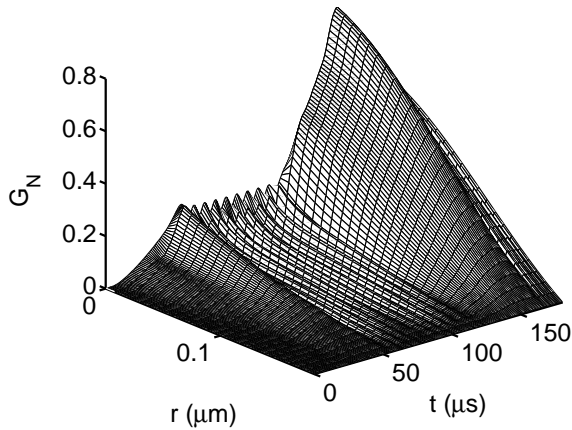}
\end{center}
\caption{
  Normal field $G_N(r)$ as a function of time and distance for the
  same calculation illustrated in Fig.~\ref{fig3}. We identify these
  atoms as the burst atoms of Donley {\it et
    al.}~\protect\cite{donley}. The spread in $r$-space is a measure
  of the energy of these noncondensate atoms.}
\label{fig4}
\end{figure}

The growth of the noncondensate component can been seen in
Fig.~\ref{fig4}.  The function $G_N(r=0)$ represents the density of
noncondensate atoms, which can be seen to oscillate out of phase with
the atomic condensate during $t_{\rm evolve}$. It is remarkable that
in spite of the fact that the Ramsey fringes occur at the molecular
binding energy with significant visibility, the population of the
molecular condensate remains small. A much larger fraction is
converted into strongly correlated atom pairs, encapsulated by the
normal and anomalous densities. Interestingly, the anomalous density
is the same aspect of the field theory which accounts for Cooper
pairing in a nonideal Fermi gas and gives rise to superfluidity at
temperatures below the critical value.  The explanation for the
observed growth of the pairing field rather than the molecular
condensate is due to the close proximity of the bound state to
threshold. The range of the molecular bound state stretches in this
case to very large internuclear distances, something which has much
more overlap with the delocalized $G_A$ pairing field than with the
localized closed channel state.

\begin{figure}
\begin{center}\
  \epsfysize=60mm \epsfbox{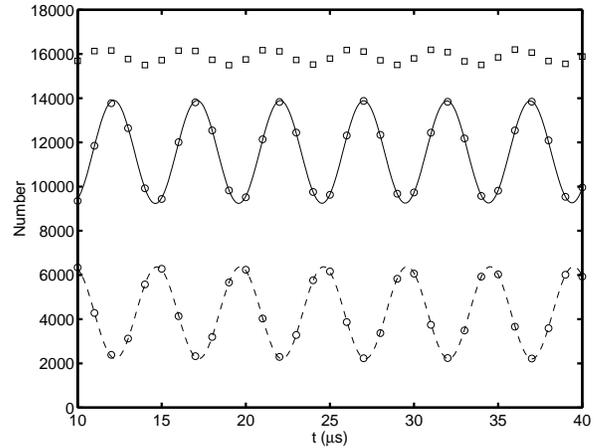}
\end{center}
\caption{
  Oscillations between the atomic condensate (solid line) and the
  normal field $G_N(0)$ (dashed line). These two numbers add up to the
  total number of recovered atoms (squares), which excludes the
  molecular component.  The calculation is performed for a mean
  density of $\langle n\rangle =3.9\times 10^{12}$ cm$^{-3}$. This
  case can be directly compared with Fig.~6 of Donley {\it et
    al.}~\protect\cite{donley}.}
\label{fig5}
\end{figure}

In Fig.~\ref{fig5} we show the population of atomic condensate and
noncondensate atoms, obtained at the end of the pulse sequence, as a
function of the evolution time $t_{\rm evolve}$. The sum of these two
numbers (squares) equals the total number of initial condensate atoms
minus twice the number of molecular condensate atoms due to particle
conservation. The frequency of the oscillations agrees with the
binding energy of the renormalized potential, given by the open
squares in Fig.~\ref{fig1}. When we compare these curves with Fig.~5
of Donley {\it et al.}~\cite{donley}, we see that we can clearly
identify the remnant observed in the experiment as the atomic
condensate component of the quantum field theory. The experimental
data closely resembles the solid curve both in offset and in
amplitude.  Similarly, the noncondensate atoms can be identified as
the burst atoms. Since $G_N(r)$ is a correlation function, it is
straightforward to determine the energy of the noncondensed atoms
which are produced. This manifests as the spatial decay rate of the
correlation function in the $r$-direction, which can be converted into
an average energy.  The energy which results is comparable to the
experimentally determined energy range for the burst.  The missing
atoms are also elucidated.  Note that there is a large time interval
between the final ramp and the time at which the atomic absorption
imaging takes place. These weakly bound molecules may decay to lower
vibrational states via a collision with a third atom, resulting in
large kinetic energies for both scattering partners. Such atoms would
not be observed.

We have repeated our calculation for a different experimental
situation~\cite{donley2} with a factor 10 larger density, and a
different time-dependence of the field. Here the number of remnant
atoms is lower than the burst atoms, so that the position of the
fringes shown in Fig.~\ref{fig5} are switched and show indications of
damping. We again get good agreement with the experiment, and the
appearance of damping of the Ramsey fringes in the theory is due to
density-dependent inhomogeneous dephasing.  This effect is due to a
relatively small shift in the oscillation frequency that shows a
strong dependence on density.  Finally we note that both theory and
experimental data exhibit a notable phase-shift between the position
of the fringes of the atomic condensate and the noncondensate atoms.

In conclusion, we have used a resonance effective field theory which
includes an accurate description of the two-body bound state and
scattering physics to describe a recent experiment at JILA. Because of
the complexity of the mean-field physics relevant to Feshbach
resonance scattering, it has not been possible previously to provide
this kind of quantitative comparison. We are able to unambiguously
identify the observed remnant and burst. The pairing field associated
with the noncondensate atoms plays a crucial role in our calculated
evolution. This pairing field is analogous to the formation of Cooper
pairs in a superfluid Fermi gas. The ability to determine the coupling
constants from known two-body rubidium physics allows us to make these
comparisons with experimental data with no adjustable parameters.

We thank J. Cooper, J. Milstein, E. Donley, N. Claussen, S. Thompson,
E. Cornell and C.  Wieman, for stimulating discussions.  Support is
acknowledged for S.K. from the U.S. Department of Energy, Office of
Basic Energy Sciences via the Chemical Sciences, Geosciences and
Biosciences Division, and for M.H. from the National Science
Foundation.


\begin{references}
\bibitem{footnote} The coupled channels binding energy has been
  obtained by using the most accurate rubidium two-body interactions
  to date~\protect\cite{kempen}.
\end{references}
\end{document}